# Knightian Analysis of the VCG Mechanism in Unrestricted Combinatorial Auctions

Alessandro Chiesa    Silvio Micali    Zeyuan Allen Zhu

MIT

March 25, 2014

**Abstract**

We consider auctions in which the players have very limited knowledge about their own valuations. Specifically, the only information that a *Knightian player* $i$ has about the profile of true valuations, $\theta^*$, consists of a set of distributions, from one of which $\theta_i^*$ has been drawn.

The VCG mechanism guarantees very high social welfare both in single- and multi-good auctions, so long as Knightian players do not select strategies that are dominated. With such Knightian players, however, we prove that the VCG mechanism guarantees very poor social welfare in unrestricted combinatorial auctions.

# 1 Introduction

In [CMZ14] we motivate the problem of mechanism design for Knightian players, and prove that (1) dominant-strategy mechanisms for single-good and multi-unit auctions cannot provide good social-welfare efficiency, but (2) the second-price and Vickrey mechanisms deliver good social-welfare performance, for these two settings, in undominated strategies.

In this report, we prove that the VCG mechanism guarantees very poor social welfare in unrestricted combinatorial auctions, if each Knightian player chooses an undominated strategy.

# 2 Model

We study *unrestricted combinatorial auctions*, where there are $n$ players and $m$ distinct goods. The set of possible allocations $\mathcal{A}$ consists of all possible partitions $A$ of $[m]$ into $1+n$ subsets, $A = (A_0, A_1, \ldots, A_n)$, where $A_0$ is the (possibly empty) set of unassigned goods and $A_i$ is the (possibly empty) set of goods assigned to player $i$.

For each player $i$, a valuation is a function mapping each possible subset of the goods to a non-negative real, and the set of all possible valuations is $\Theta_i = \{\theta_i : 2^{[m]} \to \mathbb{R}_{\geq 0} \mid \theta_i(\varnothing) = 0\}$. The profile of the players' true valuations is $\theta^* = (\theta_1^*, \ldots, \theta_n^*) \in \Theta$.

The set of possible outcomes is $\Omega \stackrel{\text{def}}{=} \mathcal{A} \times \mathbb{R}_{\geq 0}^n$. If $(A, P) \in \Omega$, we refer $P_i$ as the price charged to player $i$. We assume quasi-linear utilities. That is, the utility function $U_i$ of a player $i$ maps a valuation $\theta_i$ and an outcome $\omega = (A, P)$ to $U_i(\theta_i, \omega) \stackrel{\text{def}}{=} \theta_i(A_i) - P_i$.

If $\omega$ is a distribution over outcomes, we also denote by $U_i(\theta_i, \omega)$ the expected utility of player $i$.

## 2.1 Knightian Valuation Uncertainty

In our model, a player $i$'s sole information about $\theta^*$ consists of $\mathcal{K}_i$, a set of distributions over $\Theta_i$, from one of which $\theta_i^*$ has been drawn. (The true valuations are uncorrelated.) That is, $\mathcal{K}_i$ is $i$'s sole (and private) information about his own true valuation $\theta_i^*$. Furthermore, for every opponent $j$, $i$ has no information (or beliefs) about $\theta_j^*$ or $\mathcal{K}_j$.

Given that all he cares about is his expected (quasi-linear) utility, a player $i$ may 'collapse' each distribution $D_i \in \mathcal{K}_i$ to its expectation $\mathbb{E}_{\theta_i \sim D_i}[\theta_i]$.[1] Therefore, for unrestricted combinatorial auctions, a *mathematically equivalent* formulation of the Knightian valuation model is the following:

**Definition 2.1** (Knightian valuation model). *For each player $i$, $i$'s sole information about $\theta^*$ is a set $K_i$, the* candidate (valuation) set *of $i$, such that $\theta_i^* \in K_i \subset \Theta_i$.*

*We refer to an element of $K_i$ as a* candidate valuation.

In Knightian valuation model, a mechanism's performance will of course depend on the inaccuracy of the players' candidate sets, which we measure as follows.

**Definition 2.2.** *The candidate set $K_i$ of a player $i$ is* (at most) $\delta$-approximate *if, for each subset $S \subseteq [m]$, letting $K_i(S) \stackrel{\text{def}}{=} \{\theta_i(S) \mid \theta_i \in K_i\}$, $\sup K_i(S) - \inf K_i(S) \leq \delta$.*

*An auction is* (at most) $\delta$-approximate *if each $K_i$ is $\delta$-approximate.*

## 2.2 Social Welfare, Mechanisms, and Knightian Dominance

**Social welfare.** The social welfare of an allocation $A = (A_0, A_1, \ldots, A_n)$, $\text{SW}(A)$, is defined to be $\sum_i \theta_i^*(A_i)$; and the maximum social welfare, MSW, is defined to be $\max_{A \in \mathcal{A}} \text{SW}(A)$. (That is, SW and MSW continue to be defined relative to the players' true valuations $\theta_i^*$, whether or not the players know them exactly.)

More generally, the social welfare of an allocation $A$ relative to a valuation profile $\theta$, $\text{SW}(\theta, A)$, is $\sum_i \theta_i(A_i)$; and the maximum social welfare relative to $\theta$, $\text{MSW}(\theta)$, is $\max_{A \in \mathcal{A}} \text{SW}(\theta, A)$. Thus, $\text{SW}(A) = \text{SW}(\theta^*, A)$ and $\text{MSW} = \text{MSW}(\theta^*)$.

**Mechanisms and strategies.** A mechanism $M$ specifies, for each player $i$, a set $S_i$. We interchangeably refer to each member of $S_i$ as a pure *strategy/action/report* of $i$, and similarly, a member of $\Delta(S_i)$ a mixed strategy/action/report of $i$.

After each player $i$, simultaneously with his opponents, reports a strategy $s_i$ in $S_i$, $M$ maps the reported strategy profile $s$ to an outcome $M(s) \in \Omega$.

If $M$ is probabilistic, then $M(s) \in \Delta(\Omega)$. Thus, as per our notation, $U_i(\theta_i, M(s)) \stackrel{\text{def}}{=}$

---

[1]Whatever the auction mechanism used, this equivalence holds for any auction where each $\Theta_i$ is a *convex* set. In particular, this includes unrestricted combinatorial auctions of $m$ distinct goods.

$\mathbb{E}_{\omega \sim M(s)}[U_i(\theta_i, \omega)]$ for each player $i$.

Note that $S_i = \Theta_i$ for the direct mechanisms in the classical setting.

**The VCG mechanism.** In our auctions, the VCG mechanism, denoted VCG, maps a profile of valuations $\theta \in \Theta_1 \times \cdots \times \Theta_n$, to an outcome $(A, P)$, where

$A \in \arg\max_{A \in \mathcal{A}} \text{SW}(\theta, A)$ and, for each player $i$, $P_i = \text{MSW}(\theta_{-i}) - \sum_{j \neq i} \theta(A_i)$. Ties can be broken in any way and, if probabilistically (e.g., at random), then we extend the computation of the expected utility to also include this random choice.

**Knightian undominated strategies.** Given a mechanism $M$, a pure strategy $s_i$ of a player $i$ with a candidate set $K_i$ is *(weakly) undominated*, in symbols $s_i \in \text{UD}_i(K_i)$, if $i$ does not have another (possibly mixed) strategy $\sigma_i$ such that

(1) $\forall \theta_i \in K_i \; \forall s_{-i} \in S_{-i} \quad \mathbb{E} U_i(\theta_i, M(\sigma_i, s_{-i})) \geq U_i(\theta_i, M(s_i, s_{-i}))$, and

(2) $\exists \theta_i \in K_i \; \exists s_{-i} \in S_{-i} \quad \mathbb{E} U_i(\theta_i, M(\sigma_i, s_{-i})) > U_i(\theta_i, M(s_i, s_{-i}))$.

If $K$ is a product or a profile of candidate sets, that is, if $K = (K_1, \ldots, K_n)$ or $K = K_1 \times \cdots \times K_n$, then $\text{UD}(K) \stackrel{\text{def}}{=} \text{UD}_1(K_1) \times \cdots \times \text{UD}_n(K_n)$.

Note that the above notion of an undominated strategy is a natural extension of its classical counterpart, but other extensions are possible.

## 3 Result

To prove the inadequacy of the VCG in undominated strategies for unrestricted combinatorial auctions, it suffices to consider the case where there are exactly $n = 2$ players.[2] We show that in $\delta$-approximate combinatorial auctions with 2 players and $m$ goods, the VCG mechanism cannot, in undominated strategies, guarantee social welfare greater than $\text{MSW} - (2^m - 3)\delta$:

---

[2] If there are more than 2 players, one can always assume that the players other than 1 and 2 value 0 (and report 0 for) every subset of the goods.

**Theorem 1.** *In a combinatorial Knightian VCG auction with 2 players and $m$ goods, there exist products of $\delta$-approximate candidate sets $K = K_1 \times K_2$ and profiles $(v_1, v_2) \in \mathsf{UD}(K)$, such that*

$$\forall \theta \in K_1 \times K_2 \quad \mathsf{SW}\big(\theta, \mathsf{VCG}(v_1, v_2)\big) \leq \mathsf{MSW}(\theta) - (2^m - 3)\delta \ .$$

**Proof.** Let $\pi_1, \ldots, \pi_{2^m-1}$ be any permutation of all non-empty subsets of $[m]$ such that, whenever $j < k$, $\pi_j \not\supseteq \pi_k$.[3] We set $\pi_{2^m} \stackrel{\text{def}}{=} \pi_1$, and denote by $\overline{S}$ the complement of a subset $S$: that is, $\overline{S} \stackrel{\text{def}}{=} [m] \setminus S$.

We begin by choosing a highly-deviating strategy for player 1, and argue that it is undominated. Specifically, choose arbitrarily a real number $x$ larger than $\delta$, and then choose a candidate set $K_1$ and a strategy (i.e., a valuation) $v_1$ as follows:

$$K_1 \stackrel{\text{def}}{=} \left\{ \theta_1 \in \Theta_1 \,\middle|\, \forall \text{ non-empty } S \subseteq [m],\ \theta_1(S) \in [x - \delta/2, x + \delta/2] \right\} \text{ and}$$

$$v_1(\pi_i) \stackrel{\text{def}}{=} x + (i-1)\delta \quad \forall i \in \{1, \ldots, 2^m - 1\} \ .$$

Note that $v_1 \notin K_1$. (Indeed, $v_1(\pi_i) \in K_1(\pi_1)$ only for $i = 1$.)

We now prove that the strategy $v_1$ is undominated. More precisely,

**Claim 3.1.** $v_1 \in \mathsf{UD}_1(K_1)$.

*Proof.* We proceed by contradiction. Assume that $v_1$ is (weakly) dominated by a strategy $v_1' \neq v_1$ relative to the candidate set $K_1$. There are two cases to consider: $v_1'$ is pure and $v_i'$ is mixed.

### Proof when $v_1'$ is pure

We contradict the assumption that $v_1'$ dominates $v_1$ relative to $K_1$ by exhibiting a valuation $\theta_1 \in K_1$ and a 'witness' strategy $v_2$ for player 2 such that

$$U_1(\theta_1, \mathsf{VCG}(v_1, v_2)) > U_1(\theta_1, \mathsf{VCG}(v_1', v_2)) \ . \tag{3.1}$$

The first step is to distinguish between the following three cases (at least one of

---

[3]In particular, we can order the subsets of $[m]$ by increasing cardinality, and lexicographically within a given cardinality: that is, when $m = 3$, $\{1\}, \{2\}, \{3\}, \{1,2\}, \{1,3\}, \{2,3\}, \{1,2,3\}$.

them always holds[4]):

(1) $\exists i \in \{1, \ldots, 2^m - 1\}$, $v_1(\pi_{i+1}) - v_1(\pi_i) > \max_{T \subseteq \pi_{i+1}} v_1'(T) - \max_{T \subseteq \pi_i} v_1'(T)$

(2) $v_1(\pi_1) > \max_{T \subseteq \pi_1} v_1'(T)$

(3) $v_1(\pi_1) < \max_{T \subseteq \pi_1} v_1'(T)$

**Case (1).** Suppose that case (1) holds. Then there exists some $i \in \{1, \ldots, 2^m - 1\}$ such that

$$\exists \Delta \quad v_1(\pi_{i+1}) - v_1(\pi_i) > \Delta > \max_{T \subseteq \pi_{i+1}} v_1'(T) - \max_{T \subseteq \pi_i} v_1'(T) \ . \tag{3.2}$$

From here we distinguish two additional sub-cases:

**Sub-Case (1.1).** Inequality (3.2) holds for $i \notin \{2^m - 2, 2^m - 1\}$.

In this case neither $\pi_i$ nor $\pi_{i+1}$ is $\pi_{2^m-1} = [m]$. In addition, we can assume that $\Delta > 0$ since $v_1(\pi_{i+1}) - v_1(\pi_i) = \delta > 0$.

To show (3.1) we define $v_2$ and $\theta_1 \in K_1$ as follows. Letting $H$ be a huge number (e.g., much higher than $v_1(S)$ and $v_1'(S)$ for any subset $S$ of the goods), we define

$v_2:$ $v_2(\overline{\pi_{i+1}}) = H - \Delta$, $v_2(\overline{\pi_i}) = H$, and $v_2(S) = 0$ for all other non-empty subsets $S \subseteq [m]$;

$\theta_1:$ $\theta_1(\pi_{i+1}) = x + \delta/2$ and $\theta_1(S) = x - \delta/2$ for all other non-empty subsets $S \subseteq [m]$.

For our choices of $\theta_1, v_1, v_1'$ and $v_2$ we now argue that:

$$U_1(\theta_1, \mathsf{VCG}(v_1, v_2)) = (x + \delta/2) - \Delta \tag{3.3}$$

$$U_1(\theta_1, \mathsf{VCG}(v_1', v_2)) = (x - \delta/2) - 0 \ . \tag{3.4}$$

*Proof of (3.3)* To prove (3.3) it suffices to show that, in the outcome $\mathsf{VCG}(v_1, v_2)$, the allocation is $(\pi_{i+1}, \overline{\pi_{i+1}})$ and player 1's price is $\Delta$. Indeed, because $H$ was chosen to be sufficiently large, the only allocations we should consider are $(T, \overline{\pi_{i+1}})$ and $(T', \overline{\pi_i})$ where $T \subseteq \pi_{i+1}$ and $T' \subseteq \pi_i$. By construction $\pi_{i+1}$ maximizes $v_1(T)$ among all $T \subseteq \pi_{i+1}$, and $\pi_i$ maximizes $v_1(T)$ among all $T \subseteq \pi_i$; in particular, the only two possible allocations are $(\pi_i, \overline{\pi_i})$ and $(\pi_{i+1}, \overline{\pi_{i+1}})$. Because $v_1(\pi_{i+1}) - v_1(\pi_i) > \Delta = v_2(\overline{\pi_i}) - v_2(\overline{\pi_{i+1}})$ according to (3.2), the allocation chosen must be $(\pi_{i+1}, \overline{\pi_{i+1}})$. As for the price: player 2 is allocated $\overline{\pi_{i+1}}$ but, if player 1 did not exist, player 2 would be allocated $\overline{\pi_i}$, and gain $\Delta$ in utility; thus player 1's price is indeed $\Delta$.

---

[4]Indeed, if (1) is not satisfied, then summing up all its inequalities we get $v_1(\pi_i) = \max_{T \subseteq \pi_i} v_1'(T) + c$ for some common constant $c$. However, if neither (2) nor (3) is satisfied, then $c$ must be 0, and we get $v_1(\pi_i) = \max_{T \subseteq \pi_i} v_1'(T)$ for all $i$. Since $T$ enumerates over all possible subsets of $[m]$, this further implies that $v_1(T) = v_1'(T)$ for all $T$, which contradicts $v_1 \neq v_1'$.

*Proof of (3.4)* To prove (3.4) it suffices to show that, in the outcome $\mathsf{VCG}(v_1', v_2)$, the allocation is $(T, \overline{\pi_i})$, for some $T \subseteq \pi_i$ and $T \neq \pi_{i+1}$, and player 1's price is 0. As before, because $H$ was chosen to be sufficiently large, the only allocations we should consider are $(T, \overline{\pi_{i+1}})$ and $(T', \overline{\pi_i})$ where $T \subseteq \pi_{i+1}$ and $T' \subseteq \pi_i$. Using (3.2) again, we get

$$v_2(\overline{\pi_i}) - v_2(\overline{\pi_{i+1}}) = \Delta > \max_{T \subseteq \pi_{i+1}} v_1'(T) - \max_{T \subseteq \pi_i} v_1'(T)$$

which further implies that the allocation must be $(T, \overline{\pi_i})$. As for the price: player 2 is allocated $\overline{\pi_i}$ and, if player 1 did not exist, player 2 would still be allocated $\overline{\pi_i}$; thus player 1's price is indeed 0. In addition, we must have $T \neq \pi_{i+1}$: otherwise the allocation would be $(\pi_{i+1}, \overline{\pi_i})$, implying that $\pi_{i+1} \subsetneq \pi_i$, contradicting our choice of $\pi$.

By (3.2) and the construction of $v_1$, it is clear that $\delta = v_1(\pi_{i+1}) - v_1(\pi_i) > \Delta$. Accordingly, utility (3.3) is greater than utility (3.4). That is, inequality (3.1) holds in Sub-Case (1.1).

**Sub-Case (1.2).** Inequality (3.2) holds for $i \in \{2^m - 2, 2^m - 1\}$.

The proof that inequality (3.1) holds in sub-case 1.2 is similar to that of sub-case 1.1. The only difference is that, since this is a boundary case, one must pay attention to the fact that either $\overline{\pi_i}$ or $\overline{\pi_{i+1}}$ may be the empty set. That is, we cannot set $v_2(\overline{\pi_{i+1}}) = H - \Delta$ or $v_2(\overline{\pi_i}) = H$, for an arbitrarily large $H$, because $v_2(\emptyset)$ must be zero. Instead, one must choose the value of $H$ to precisely coincide with $v_2(\emptyset) = 0$.

**Case (2).** Suppose that case (2) holds. Then,

$$\exists \Delta > 0 \quad v_1(\pi_1) > \Delta > \max_{T \subseteq \pi_1} v_1'(T) . \tag{3.5}$$

To show (3.1) we define $v_2$ and $\theta_1 \in K_1$ as follows. Letting $H$ be a huge number

(e.g., much higher than $v_1(S)$ and $v'_1(S)$ for any subset $S$ of the goods), we define

$v_2:$ $v_2(\overline{\pi_1}) = H$, $v_2([m]) = H + \Delta$, and $v_2(S) = 0$ for all other non-empty subsets $S \subseteq [m]$;

$\theta_1:$ $\theta_1(S) = x + \delta/2$ for all non-empty $S \subseteq [m]$.

For our choices of $\theta_1, v_1, v'_1,$ and $v_2$ we now argue that:

$$U_1(\theta_1, \mathsf{VCG}(v_1, v_2)) = (x + \delta/2) - \Delta \tag{3.6}$$

$$U_1(\theta_1, \mathsf{VCG}(v'_1, v_2)) = 0 \ . \tag{3.7}$$

*Proof of (3.6)* To prove (3.6) it suffices to show that, in the outcome $\mathsf{VCG}(v_1, v_2)$, the allocation is $(\pi_1, \overline{\pi_1})$ and player 1's price is $\Delta$. Indeed, because $H$ was chosen to be sufficiently large, the only allocations we should consider are $(T, \overline{\pi_1})$ and $(\varnothing, [m])$ where $T \subseteq \pi_i$. By construction $\pi_1$ maximizes $v_1(T)$ among all $T \subseteq \pi_i$, so the only two possible allocations are $(\pi_1, \overline{\pi_1})$ and $(\varnothing, [m])$. Because $v_1(\pi_1) + v_2(\overline{\pi_1}) = v_1(\pi_1) + H > H + \Delta = v_2([m])$ according to (3.5), the allocation chosen must be $(\pi_i, \overline{\pi_i})$. As for the price: player 2 is allocated $\overline{\pi_1}$ but, if player 1 did not exist, player 2 would be allocated $[m]$, and gain $\Delta$ in utility; thus player 1's price is indeed $\Delta$.

*Proof of (3.7)* To prove (3.7) it suffices to show that, in the outcome $\mathsf{VCG}(v'_1, v_2)$, the allocation is $(\varnothing, [m])$. As before, because $H$ was chosen to be sufficiently large, the only allocations we should consider are $(T, \overline{\pi_1})$ and $(\varnothing, [m])$ where $T \subseteq \pi_i$. This time by relying on the fact that $v_2([m]) = H + \Delta > H + \max_{T \subseteq \pi_1} v'_1(T)$ from (3.5), we deduce that the allocation is in fact $(\varnothing, [m])$.

By (3.5), $x = v_1(\pi_1) > \Delta$. Therefore, utility (3.6) is greater than utility (3.7), yielding (3.1). That is, inequality (3.1) holds in Case (2).

**Case (3).** Suppose that case (3) holds. Then,

$$\exists \Delta > 0 \quad v_1(\pi_1) < \Delta < \max_{T \subseteq \pi_1} v'_1(T) \ .$$

The proof that inequality (3.1) holds in Case (3) is totally symmetrical to that of Case (2).

In all cases above we have shown that (3.1) holds, thus finishing the proof of Claim 3.1 when $v_1'$ is a pure strategy.

### Proof when $v_1'$ is mixed.

Assuming that $v_1$ is dominated by a mixed strategy $v_1'$, we reach a contradiction by proving that inequality (3.1) holds in expectation. The ideas of the proof are the same, although the analysis must now be applied to all the pure strategies in the support of $v_1'$, rather than to $v_1'$ itself, which makes the argument notationally more involved.

This completes the proof of Claim 3.1. $\square$

Having constructed $v_1 \in \mathsf{UD}_1(K_1)$, we continue the proof of Theorem 1 letting:

$$v_2(S) \stackrel{\text{def}}{=} \begin{cases} (2^m - i - 1.5)\delta & \text{if } S = \overline{\pi_i} \text{ for some } i \in \{1, \ldots, 2^m - 2\} \\ x + (2^m - 2.5)\delta & \text{if } S = [m] \end{cases},$$

$$K_2 \stackrel{\text{def}}{=} \left\{ \theta_2 \in \Theta_2 \,\middle|\, \forall i \in \{1, \ldots, 2^m - 1\},\, \theta_2(\pi_i) \in \left[v_2(\pi_i), v_2(\pi_i) + \delta\right] \right\}.$$

Note that, by construction, $v_2 \in K_2$, which easily implies the following

**Claim 3.2.** $v_2 \in \mathsf{UD}_2(K_2)$.

*Proof.* Suppose, by the way of contradiction, that there exists a strategy $v_2'$, different from $v_2$, that (weakly) dominates $v_2$ relative to $K_2$. Then, by (the first property of) the definition of Knightian dominance, we have

$$\forall \theta_2 \in K_2 \;\forall v_1 \in \Theta_1 \quad \mathbb{E}U_2(\theta_2, \mathsf{VCG}(v_1, v_2)) \leq \mathbb{E}U_2(\theta_2, \mathsf{VCG}(v_1, v_2')) \,.$$

Since $v_2 \in K_2$, we can choose $\theta_2 = v_2$ in the above inequality, getting

$$\forall v_1 \in \Theta_1 \quad \mathbb{E}U_2(v_2, \mathsf{VCG}(v_1, v_2)) \leq \mathbb{E}U_2(v_2, \mathsf{VCG}(v_1, v_2')) \,.$$

However, the fact that the VCG mechanism is dominant-strategy-truthful in the exact-valuation world tells us that

$$\forall v_1 \in \Theta_1 \quad \mathbb{E}U_2(v_2, \mathsf{VCG}(v_1, v_2)) \geq \mathbb{E}U_2(v_2, \mathsf{VCG}(v_1, v_2')) \,.$$

This implies that the two different strategies $v_2$ and $v_2'$ are *equivalent* in player 2's perspective! However, since $v_2$ is strictly monotone —i.e., $v_2(S) < v_2(T)$ for any $S \subsetneq T$— no different but equivalent strategy exists for $v_2$. Indeed, for any $v_2' \neq v_2$, one can always find a strategy $v_1 \in \Theta_1$, so that player 2's utilities in the outcomes $\mathsf{VCG}(v_1, v_2)$ and $\mathsf{VCG}(v_1, v_2')$ are different. We thus have reached a contradiction and proved Claim 3.2. $\square$

Having specified $K_1, v_1, K_2$, and $v_2$, all we have left is analyzing the social welfare performance.

Let us first compute the allocation of the outcome $\mathsf{VCG}(v_1, v_2)$. The only allocations to consider are $(\pi_{2^m-1}, \varnothing)$, $(\varnothing, \pi_{2^m-1})$, and $(\pi_i, \overline{\pi_i})$, for some index $i \in \{1, \ldots, 2^m - 2\}$. (In principle, one may also consider allocations where some goods remain unallocated. However, since $v_1$ and $v_2$ are strictly monotone —that is, $v_j(S) < v_j(T)$ for all $S \subsetneq T$ and all $j \in \{1, 2\}$— all goods must be allocated in the outcome of $\mathsf{VCG}(v_1, v_2)$.)

Now we compare the social welfare relative to $(v_1, v_2)$ for such allocations:

$$v_1(\pi_{2^m-1}) + v_2(\varnothing) = (x + (2^m - 2)\delta) + 0 = x + (2^m - 2)\delta ,$$
$$v_1(\varnothing) + v_2(\pi_{2^m-1}) = 0 + (x + (2^m - 2.5)\delta) = x + (2^m - 2.5)\delta , \text{ and}$$
$$v_1(\pi_i) + v_2(\overline{\pi_i}) = (x + (i - 1)\delta) + (2^m - i - 1.5)\delta = x + (2^m - 2.5)\delta .$$

Thus, in the outcome $\mathsf{VCG}(v_1, v_2)$ the allocation is $(\pi_{2^m-1}, \varnothing)$. Hence, the social welfare is

$$\mathrm{SW}\big((\theta_1, \theta_2), \mathsf{VCG}(v_1, v_2)\big) = \theta_1(\pi_{2^m-1}) .$$

On the other hand, the maximum social welfare is

$$\mathrm{MSW}(\theta_1, \theta_2) \geq \theta_2(\pi_{2^m-1}) .$$

Now notice that for all $\theta \in K$, we have

$$\mathrm{MSW}(\theta) - \mathrm{SW}\big(\theta, \mathsf{VCG}(v_1, v_2)\big) \geq \theta_2(\pi_{2^m-1}) - \theta_1(\pi_{2^m-1}) \geq (x + (2^m - 2.5)\delta) - (x + \delta/2) = (2^m - 3)\delta .$$

This concludes our proof of Theorem 1. ∎

2014. to appear.